# A Satisfiability Algorithm for $\mathbf{AC}^0$

Russell Impagliazzo [*]    William Matthews [†]

Ramamohan Paturi [†]

Department of Computer Science and Engineering
University of California, San Diego
La Jolla, CA 92093-0404, USA

January 2011

## Abstract

We consider the problem of efficiently enumerating the satisfying assignments to $\mathbf{AC}^0$ circuits. We give a zero-error randomized algorithm which takes an $\mathbf{AC}^0$ circuit as input and constructs a set of restrictions which partition $\{0,1\}^n$ so that under each restriction the value of the circuit is constant. Let $d$ denote the depth of the circuit and $cn$ denote the number of gates. This algorithm runs in time $|C|2^{n(1-\mu_{c,d})}$ where $|C|$ is the size of the circuit for $\mu_{c,d} \geq 1/O\left(\lg c + d \lg d\right)^{d-1}$ with probability at least $1 - 2^{-n}$.

As a result, we get improved exponential time algorithms for $\mathbf{AC}^0$ circuit satisfiability and for counting solutions. In addition, we get an improved bound on the correlation of $\mathbf{AC}^0$ circuits with parity.

As an important component of our analysis, we extend the Håstad Switching Lemma to handle multiple $k$-CNFs and $k$-DNFs.

---

[*]Work supported by the Ellentuck Fund, Friends of the Institute for Advanced Study, and NSF grants DMS-0835373 and CCF-0832797 subcontract No. 00001583. Any opinions, findings and conclusions or recommendations expressed in this material are those of the author and do not necessarily reflect the views of the National Science Foundation, Ellentuck Fund, or Friends of the Institute for Advanced Study.

[†]This research is supported by NSF grant CCF-0947262 from the Division of Computing and Communication Foundations. Any opinions, findings and conclusions or recommendations expressed in this material are those of the authors and do not necessarily reflect the views of the National Science Foundation.

# 1 Introduction

The Circuit Satisfiability problem, deciding whether a Boolean circuit has an assignment where it evaluates to true, is in many ways the canonical **NP**-complete problem. It is not only important from a theoretical point of view, but pragmatically, since search problems directly reduce to Circuit Satisfiability without increasing the size of the search space. In contrast, reducing problems such as planning or model-checking to the special cases of CNF Satisfiability or 3-SAT can increase the number of variables dramatically. On the other hand, CNF-SAT algorithms (especially for small width clauses such as $k$-CNFs) have both been analyzed theoretically and implemented empirically with astounding success, even when this overhead is taken into account. For some domains of structured CNF formulas that arise in many applications, state-of-the-art SAT-solvers can handle up to hundreds of thousands of variables and millions of clauses [11]. Meanwhile, very few results are known either theoretically or empirically about the difficulty of Circuit Satisfiability.

This raises the question: for which classes of circuits are there non-trivial Satisfiability algorithms? More precisely, for an algorithm that decides satisfiability for circuits in a class $\mathcal{C}$, we express its worst-case running time as $|C|2^{n(1-\mu)}$ where $C$ is a circuit with $n$ inputs. We say that $\mu$ is the savings over exhaustive search and write it in terms of $n$ and the parameters of the class $\mathcal{C}$, for example, the ratio of gates or wires to the number of inputs, and depth. The larger $\mu$ is, the more non-trivial the algorithm can be considered. When can we exploit the structural properties of $\mathcal{C}$ to obtain savings over exhaustive search? What are the best savings for various circuit classes? How is the expressive power of a circuit class related to the amount of savings for its satisfiability problem?

Williams' recent work [20, 21] has given an additional motivation to look at improved algorithms for Circuit Satisfiability problems. He shows that Circuit Satisfiablity algorithms for a class of circuits with even a very small improvement ($\mu = \omega(\log n/n)$) over exhaustive search imply circuit lower bounds for that class. Thus, he formally proves a connection showing that the same understanding of the limitations of a circuit class is needed both for improved algorithms and to prove lower bounds.

Among the satisfiability problems for various circuit classes, the $k$-SAT problem has attracted the attention of several researchers since Monien and Speckenmeyer [12] first showed that it can be computed in less than $2^n$ time in the worst-case, that is, in time $|F|2^{n(1-\mu_k)}$ for $\mu_k > 0$ where $F$ is a $k$-CNF. Researchers have obtained a series of improvements for $\mu_k$, with particular attention to the case $k = 3$. Currently, the best known savings for 3-SAT is about 0.614 for randomized algorithms [14, 9] and about 0.4157 for deterministic algorithms [18, 13]. There are a suprisingly wide variety of state-of-the-art algorithms for $k$-SAT, but they all achieve savings $\mu_k$ of the form $\Theta(1/k)$ [15, 18].

Much less is known about the nature of savings for satisfiability problems for more expressive circuit classes. Schuler [19] has shown that the satisfiability of an $m$-clause CNF $F$ can be determined in time $|F|2^{n(1-\frac{1}{1+\lg m})}$. Using a bit more careful analysis, Calabro et al. [3] improved this from savings $1/O(\lg m)$ to $1/O\left(\lg \frac{m}{n}\right)$, giving constant savings when $m = O(n)$.

Subsequently, Calabro et al. [4] have considered the question of satisfiability of bounded-depth unbounded fan-in circuits over standard basis (**AC**$^0$ circuits) and shown that the satisfiability of **AC**$^0$ circuits $C$ of size $cn$ and depth $d$ can be decided in time $|C|2^{n(1-\mu)}$ where $\mu \geq 1/O(c^{2^{d-2}-1} \lg^{3 \cdot 2^{d-2}-2} c)$. Santhanam [17] has considered $cn$-size formulas $F$ with no depth restriction and showed that the satisfiability problem for such formulas can be solved in time $|F|2^{n\left(1-\frac{1}{poly(c)}\right)}$. More recently, Williams [21] has shown that the satisfiability of **ACC** circuits $C$ can be solved in time $|C|2^{n-\Theta(n^{2-\Theta(d)})}$.

In this paper, we return to $\mathbf{AC}^0$ circuits and seek further improvement in the savings for the satisfiability algorithm for $\mathbf{AC}^0$ circuits of size $cn$ and depth $d$. While the algorithm in [4] obtains savings in terms of $c$ and $d$ independent of $n$, its double exponentially small savings diminish rapidly to zero for $d > 2$ when $c$ grows as a function of $n$. While Williams' algorithm [21] provides nontrivial savings even when $c$ grows sufficiently large as a function of $n$, its savings decreases with $n$ even when $c$ and $d$ are constants. Furthermore, it leaves open the question whether one can obtain better savings for $\mathbf{AC}^0$ circuits.

Another independent motivation is the natural connection to proving lower bounds. Paturi et al. [15] have observed that the analysis that led to an improved upper bound for $k$-SAT can also be used to prove lower bounds for depth-3 $\mathbf{AC}^0$ circuits. Using this approach, they obtain a tight lower bound of $\Omega(n^{\frac{1}{4}} 2^{\sqrt{n}})$ on the number of gates required to compute the parity function with depth-3 circuits. Subsequently, [14] proposed a resolution-based $k$-SAT algorithm and obtained an improved savings (by a constant factor) using a sophisticated analysis. Using the same analysis, they construct a fairly simple function (checking whether the input binary string is a codeword of a certain code) which requires at least $2^{1.282\sqrt{n}}$ size for any depth-3 $\mathbf{AC}^0$ circuit. This is the best-known lower bound for depth-3 circuits for any function. Although depth reduction techniques based on Switching Lemma [5, 16, 1] and especially the top-down technique of [7] for depth-3 circuits prove lower bounds that are close to $2^{\sqrt{n}}$, it is not clear that these techniques by themselves would yield a lower bound of $2^{c\sqrt{n}}$ for $c > 1$.

It has been a long open problem to prove $2^{\omega(n^{\frac{1}{d-1}})}$ size lower bounds for $\mathbf{AC}^0$ circuits of depth $d$. Since the breakthrough results of Yao and Håstad in the mid 1980's, there have been only modest improvements [22, 5] and only for for depth-3 circuits [7, 14]. It is tantalizing to prove better lower bounds by exploiting the connection between $\mathbf{AC}^0$ satisfiability upper bounds and lower bounds. Unfortunately, we do not have any ideas for proving strong enough upper bounds for $\mathbf{AC}^0$ satisfiability that would imply better lower bounds. Our modest goal is to obtain a satisfiability algorithm for $\mathbf{AC}^0$ with savings sufficient enough to imply the some of the best-known lower bounds.

## 1.1 Our Results and Techniques

Our main result is an algorithm for $\mathbf{AC}^0$ satisfiability with the best known savings.

**Theorem 1.1.** *There is a Las Vegas algorithm for deciding the satisfiability of circuits with $cn$ gates and depth $d$ whose expected time has savings at least $\frac{1}{O(\lg c + d \lg d)^{d-1}}$*

The above algorithm immediately follows from the existence of an algorithm that enumerates all satisfying assignments by partitioning the space into sub-cubes where the circuit is constant.

**Theorem 1.2.** *There exists a Las Vegas algorithm which, on input a $cn$ gate, depth $d$ circuit $C$ in $n$ variables, produces a set of restrictions $\{\rho_i\}_i$ which partition $\{0,1\}^n$ and such that for each $i$, $C|_{\rho_i}$ is constant. The expected time and number of restrictions are both $poly(n)|C|2^{n(1-\mu_{c,d})}$, where the savings $\mu_{c,d}$ is at least $\frac{1}{O(\lg c + d \lg d)^{d-1}}$.*

This result implies the known, almost tight, lower bounds for depth-size tradeoffs for computing parity. This is not surprising, because we use a version of the Håstad Switching Lemma that was used to prove these lower bounds. However, our results also imply new, very tight, bounds on how well constant-depth circuits can approximate the parity. As another corollary to Theorem 1.2, we obtain the following bounds on correlation between $\mathbf{AC}^0$ circuits and the parity function. Recently and independently from this work, Håstad [6] achieved a similar bound on correlation (although our result is better for small values of $m$).

**Theorem 1.3.** *The correlation of parity with any* $\mathbf{AC}^0$ *circuit of size cn and depth d is at most* $2^{-\mu_{c,d}n} = 2^{-n/O(\log c + d \log d)^{d-1}}$.

Some particularly interesting special cases are:

1. For linear sized families of circuits, with $c, d$ constant, we obtain a constant savings $\mu$ and a strongly exponential bound $2^{-\Theta(n)}$ on the correlation with parity.

   ([4] also gives constant savings for Satisfiability, but our constant improves on theirs more than exponentially and also holds for counting.)

2. We obtain non-trivial savings and correlation bounds for $m$ up to $2^{O(n^{1/(d-1)})}$.

   Independently, Håstad [6] achieved a similar result for correlation. Beame, Impagliazzo, and Srinivasan [2] also give an improved algorithm and correlation bound, but only for $m$ up to a quasi-polynomial size.

3. For $k$-CNF's, we extend the known savings $\mu = \Theta(1/k)$ for Satisfiability to also include counting and enumeration of solutions.

   The previous best algorithms for counting [10] had savings $\mu \approx 2^{-k}$.

Our results seem tight in the following two ways. For covering the set of solutions by sub-cubes, we achieve the best possible $\mu$ up to a constant factor. Also, any algorithm that improved our $\mu$ more than polynomially would prove that $\mathbf{NEXP} \not\subseteq \mathbf{NC}^1$ using Williams' technique.

A key ingredient in our analysis is an extended Switching Lemma, proved in section 3.2.

**Lemma 1.4** (Extended Switching Lemma). *Let $\phi_1, \ldots, \phi_m$ be a sequence of $k$-CNFs and/or $k$-DNFs in the same $n$ variables. For any $p \leq 1/13$, let $\rho$ be a random restriction which leaves $pn$ variables unset. The probability that the decision tree for $(\phi_1, \ldots, \phi_m)|_\rho$ has a path of length $\geq s$ where each $\phi_i$ contributes at least one node to the path is at most $(13pk)^s$.*

Using the observation that any path in the decision tree for $(\phi_1, \ldots, \phi_m)|_\rho$ is also a path in the decision tree for the subset of $k$-CNFs or $k$-DNFs which contribute nodes to the path, and using a union bound, we get the following corollary.

**Corollary 1.5.** *Let $\phi_1, \ldots, \phi_m$ be a sequence of $k$-CNFs and/or $k$-DNFs in the same $n$ variables. For any $p \leq 1/13$, let $\rho$ be a random restriction which leaves $pn$ variables unset. The probability that the decision tree for $(\phi_1, \ldots, \phi_m)|_\rho$ has a path of length $\geq s$ is at most $(2^m - 1)(13pk)^s$.*

In the rest of the paper, we provide detailed algorithms and analysis to support our results.

## 2 Notation

We consider layered boolean circuits with alternating layers of AND and OR gates, and where negations are all at the input level. We refer to a circuit with $d$ layers of gates as a depth $d$ circuit. We number the layers from the output gate, layer 1, to the inputs and their negations (layer $d+1$).

For circuits on $n$ inputs, we parameterize them by $m = cn$, the maximum number of gates in a layer and $d$, the number of layers. We call such circuits $(n, m, d)$-*circuits*, where $m$ and $d$ could be functions of $n$. For technical reasons we are also interested in a slight variant of $(n, m, d)$-circuits where we only require that each gate at level $d$ has fan-in bounded by $k$ (rather than limiting the number of gates at level $d$), for some $k$. All other layers are still required to have at most $m$ gates. We refer to these circuits as $(n, m, d, k)$-*circuits*.

A restriction $\rho$ on a set of variables $V$ is a map $\rho : V \to \{0, 1, *\}$. We say that the variables $v$ where $\rho(v) = *$ are *unset*. For Boolean function $f$ on all $n$ inputs, $f|_\rho$ is the restricted function on the unset variables. For a circuit $C$, $C|_\rho$ is the restricted circuit on the unset variables.

We say that a set of functions $\phi_1, \ldots, \phi_m : \{0, 1\}^n \to \{0, 1\}$ partitions $\{0, 1\}^n$ if for every $x \in \{0, 1\}^n$ there exists exactly one $i$ such that $\phi_i(x) = 1$. The $i$'th region of the partition is the set of $x$ so that $\phi_i(x) = 1$, and we identify the region with the function $\phi_i$. In particular, we are interested in partitions defined by functions of the form $R \wedge \rho$, where $R$ is a $k$-CNF and $\rho$ is a restriction, which we will abbreviate as $\mathcal{R} = (R, \rho)$. We say that two circuits, $C$ and $D$, are equivalent in a region $\mathcal{R}$ if $\mathcal{R} \implies (C \equiv D)$.

**Definition 2.1.** Let $C$ be a circuit in $n$ variables. We say that a set $\mathcal{P} = \{(\mathcal{R}_i = (R_i, \rho_i), C_i)\}_i$ is a *partitioning for $C$* if $\{\mathcal{R}_i\}_i$ defines a set of regions which partition $\{0, 1\}^n$ and for every $i$, $C$ is equivalent to $C_i$ in the region $\mathcal{R}_i$. We say that $C_i$ is the *circuit associated with* the region $\mathcal{R}_i$.

We generalize this definition to sequences of circuits in the natural way.

Define the height of a decision tree $T$, $height(T)$, as the length of the longest path. We will view paths in decision trees as restrictions in the natural way. The *canoncial decision tree* for a CNF $\phi$, $tree(\phi)$ is constructed as follows: If any clause is empty, return 0. If there are no clauses, return 1. Otherwise, let $C$ be the first clause. Query all the variables in $C$ in order. Restrict $\phi$ by the results of these queries and recurse. The canonical decision tree for a DNF is analogous.

We also similarly define the canonical decision tree for a sequence of CNFs and/or DNFs $\Phi = (\phi_1, \ldots, \phi_\ell)$. First construct the canonical decision tree for $\phi_1$. Along each path, restrict $\phi_2, \ldots, \phi_\ell$ by the answers to queries and recurse. Label the leaves of the resulting tree with the tuples of leaves from the original trees.

When we construct the canonical decision tree for a CNF $\phi$, we group variables by the clause we are at when we query them, and say the clause *contributes* the corresponding set to the path. Similarly for any subcircuit $\psi$ of $\phi$, we regard the set of nodes contributed by all the occurrences of variables in $\psi$ as the set of nodes contributed by $\psi$.

## 3 Main Algorithm

Schuler [19] gives an algorithm for CNF-SAT with savings $O(1/\log c)$. He does this by reducing the CNF to a moderately exponential sized set of $O(\log c)$-CNFs through a case analysis, then using known $k$-SAT algorithms such as [15] to get overall savings $1/O(\log c)$. Like [4], our algorithm can be thought of as generalizing Schuler's approach to larger depths, where each step performs a case analysis to reduce a single circuit of depth $d$ to a moderate-sized collection of circuits which are all depth $d - 1$. Calabro et. al. use a complex, but local, form of Schuler's case analysis, branching on large constant sized sub-formulas of the input formula. Here, we use a more global case analysis, based on Håstad's Switching Lemma. This global treatment was in part inspired by an algorithm of Santhanam [17] for formula satisfiability, that constructs a decision tree for the formula whose paths are short on average, although there might not be a small depth decision tree for the formula.

Using a Switching Lemma to convert depth $d$ circuits to depth $d - 1$ circuits is standard. However, to achieve the claimed savings with Switching Lemmas is not at all straight-forward. Assume that we have converted the bottom two levels of our circuit to $k$-CNFs. The main idea is that after a random restriction setting all but about $n/k$ variables, with high probability, each of these sub-circuits is equivalent to a small depth decision tree. View the random restriction as first picking the set of variables to restrict, then the values. For a randomly chosen set of variables to restrict, our algorithm will perform an exhaustive search over all settings of the values. If for a certain setting, the sub-formulas all become decision trees of depth $k'$, we can write them all as $k'$-DNFs and combine with the level of $OR$ gates above. Then we can hope to get the recursive savings on these branches for depth $d-1$ circuits on $n' = \Omega(n/k)$ variables.

So there are two factors that limit our savings with this approach: Even if the formula became constant after every restriction, we cannot get savings more than $1/k$, since we use exhaustive search on $n - n/k$ variables.. On the other hand, there is a failure probability (exponentially small in $k'$) where our sub-formulas might not become small depth decision trees. For these branches, we might not get any savings. So our savings is also bounded by roughly $k'/n$, the log of the failure probability as a fraction of $n$. Since our value of $k'$ in this depth becomes the new value of $k$, it is hard to see how to get savings more than $1/\sqrt{n}$ by this type of argument, even for depth 3.

We get around this by taking a more error-tolerant approach. We don't assume that all of our sub-formulas become small depth decision trees, just most of them. This is still problematic, because the sub-formulas might be identical or close, so the events that they become small depth decision trees might be highly correlated. To handle this, we look not at the decision tree complexity of a single sub-formula but at sets of sub-formulas, where the decision tree has to compute the value of each one. If several formulas become high depth for essentially the same reason, once we've evaluated one of them, the others should become small depth, so the combined decision tree will still have relatively small depth. We give an extended switching lemma that proves that it is (almost) exponentially unlikely that there is a large set of sub-formulas who all contribute many variables to their joint decision tree. Thus, intuitively, with extremely high probability, either only a few sub-formulas remain complex, or the ones that do all involve the same moderate sized subset of variables.

Our algorithm then does a case analysis over which of the sub-formulas are in this set of mutually complex ones, and over all paths in their joint decision tree. For each, the other sub-formulas are equivalent to $k$-DNFs by definition. Because we are not insisting that all sub-formulas become small, we can pick $k$ relatively small and still have an extremely small chance of failure. $k$ affects the overhead for the case analysis (we'll have to branch on which set of at most $n/k$ of the $m$ sub-formulas stay complex, which is relatively small for $k = O(\log c)$), which is determined by the length of the joint decision tree. We can let this be a constant fraction of the remaining variables, and so get a failure probability exponentially small in $n/k$ rather than $k$.

Here, we describe the overall structure of the algorithm. In Section 3.1 we describe the algorithm for converting one level of CNFs to DNFs and vice versa, and in Section 3.2 we prove our extended switching lemma.

We construct a partition for an $(n, m, d)$-circuit into regions and associated $(n', m, d, k)$-circuits by using a technique of Schuler [19]. For the sake of exposition, we assume that the bottom level gates are $\vee$ gates. The other case can be handled in an analogous fashion. While there exist a bottom level gate $\phi$ of fan-in greater than $k$, branch on the disjunction of the first $k$ inputs of $\phi$. If the disjunction is true, we replace the gate by the smaller $k$ input disjunction. If the disjunction is false, we set the value of the $k$ variables in the disjunction. We repeat this branching process until all bottom level gates have fan-in at most $k$. This step reduces the overall savings by about $2^{-k}m/n$.

**Lemma 3.1** (Bottom Fan-in Reduction Algorithm)**.** *Let $C$ be a $(n, m, d)$-circuit and let $k \geq 1$ be a parameter. There exists an algorithm which outputs a partitioning $\mathcal{P} = \cup_{0 \leq f \leq n/k} \mathcal{P}_f$ for $C$, where the sets $\mathcal{P}_f$ are disjoint and for each $f$, $\mathcal{P}_f$ contains at most $\binom{m+f}{f}$ regions with associated $(n - fk, m, d, k)$-circuits. The algorithm runs in time $poly(n) \cdot |C| \cdot |\mathcal{P}|$.*

*Proof.* Let $C$ be a $(n, m, d)$-circuit and let $k \geq 1$ be a parameter. Let $R$ be an empty (true) $k$-CNF

and let $\rho$ be a restriction where all variables are unset. Assume that the bottom level gates are $\vee$ gates.

While there exists a bottom level gate $\phi$ with fan-in greater than $k$, let $\phi'$ denote the disjunction of the first $k$ inputs of $\phi$. Branch on $\phi'$. In the branch where $\phi'$ is true, replace $\phi$ with $\phi'$ in $C$ and replace $R$ with $R \wedge \phi'$. In the branch where $\phi'$ is false, replace $\rho$ with $\rho \wedge \neg\phi'$, viewing $\neg\phi'$ as a restriction which sets the literals in $\phi'$ to false and replace $C$ with $C|_{\neg\phi'}$.

When all bottom level gates in $C$ have fan-in at most $k$, output $(\mathcal{R} = (R, \rho), C)$.

Along any path in the computation tree of this algorithm, let $f$ denote the number of branches where $\phi'$ is false. Note that each path has at most $m$ branches where $\phi'$ is true since each such branch reduces the number of bottom level gates with fan-in greater than $k$. For each value of $f$, there are at most $\binom{m+f}{f}$ paths. At the end of each such path, $fk$ variables are set by the false branches, resulting in $(n - fk, m, d, k)$-circuits. □

Next, we repeatedly reduce the depth of $(n, m, i, k)$-circuits by one until it reaches depth 2 (either a $k$-CNF or a $k$-DNF). Let $\Phi$ be the sequence of subcircuits of an $(n, m, i, k)$-circuit at depth $i - 1$. Assume without loss of generality that the subcircuits are $k$-DNFs. The main technical ingredient for depth reduction is an algorithm which constructs a partition which allows us to transform a sequence of $k$-DNFs into a sequence of equivalent $k$-CNFs in each region, or vice versa. This algorithm will be described in detail in Section 3.1. We apply this algorithm to transform $\Phi$ into sequences of $k$-CNFs. Since the gates at level $i - 1$ change from $\vee$ to $\wedge$, they may be combined with the gates at level $i - 2$ to reduce the depth by one without increasing the number of gates at any levels $i - 2$ or higher.

**Lemma 3.2** (Depth Reduction Algorithm). *Let $C$ be a $(n, m, d, k)$-circuit and let $0 < q \leq 1/2$ be a parameter. There exists a randomized algorithm which outputs partitioning $\mathcal{P}$ for $C$ where the circuit associated with each region of $\mathcal{P}$ is a $(\frac{n}{100k}, m, d-1, k)$-circuit. With probability at least $1 - q$, $|\mathcal{P}| \leq s$ and the algorithm runs in time $poly(n) \cdot \lg \frac{1}{q} \cdot |C| \cdot s$ where $s \leq \frac{2n}{100k} \cdot 2^{n - \frac{n}{100k} + 3^{-k}m}$.*

The depth reduction algorithm follows from Lemma 3.3, given below and proved in Section 3.1.

**Lemma 3.3** (Switching Algorithm). *Let $\Phi = (\phi_1, \ldots, \phi_m)$ be a sequence of $k$-CNFs in $n$ variables and let $0 < q \leq 1/2$ be a parameter. There exists a randomized algorithm which takes $\Phi$ as input and outputs a partitioning $\mathcal{P}$ for $\Phi$ where the circuits associated with each region of $\mathcal{P}$ are $k$-DNFs in at most $\frac{n}{100k}$ variables. With probability at least $1 - q$, $|\mathcal{P}| \leq s$ and the algorithm runs in time $poly(n) \cdot \lg \frac{1}{q} \cdot |\Phi| \cdot s$ where $s \leq \frac{2n}{100k} \cdot 2^{n - \frac{n}{100k} + 3^{-k}m}$.*

The case where each $\phi_j$ is a $k$-CNF and the circuits in each region of $\mathcal{P}$ are $k$-DNFs is symmetric.

*Proof of Lemma 3.2.* Let $C$ be a $(n, m, d, k)$-circuit and let $0 < q \leq 1/2$ be a parameter. Let $\Phi = (\phi_1, \ldots, \phi_m)$ be the sequence of subcircuits at rooted at level $d - 1$ in $C$ ($k$-CNFs or $k$-DNFs). Run the Switching Algorithm (Lemma 3.3) on $\Phi$ to get $\{(\mathcal{R}_i, \Psi_i = (\psi_{i,1}, \ldots, \psi_{i,m}))\}_i$ ($\Psi_i$ will be a sequence of $k$-DNFs or $k$-CNFs). For each $i$, let $C_i$ be the circuit resulting from replacing $\phi_1, \ldots, \phi_m$ with $\psi_{i,1}, \ldots, \psi_{i,m}$ in $C$ and then combining the gates at level $d - 2$ and $d - 1$ (which will be the same type of gates). Output $(\mathcal{R}_i, C_i)$.

By Lemma 3.3, each $\Psi_i$ is a depth 2 circuit in at most $\frac{n}{100k}$ variables. Therefore, each $C_i$ will be a $(\frac{n}{100k}, m, d-1, k)$-circuit after combining the gates at levels $d-2$ and $d-1$. With probability at least $1 - q$, the algorithm of Lemma 3.3 produces $\mathcal{P}$ satisfying $|\mathcal{P}| \leq s$ and runs in time $poly(n) \cdot \lg \frac{1}{q} \cdot |\Phi| \cdot s$ where $s \leq \frac{2n}{100k} \cdot 2^{n - \frac{n}{100k} + 3^{-k}m}$. This algorithm produces a partition of the same size and increases the running time by an additive $O(|C| \cdot |\mathcal{P}|)$. □

This lemma does $d-2$ steps of depth reduction.

**Lemma 3.4** (Repeated Depth Reduction Algorithm). *Let $C$ be a $(n,m,d,k)$-circuit and let $0 < q \leq 1/2$ be a parameter. There exists a randomized algorithm which outputs a partitioning $\mathcal{P}$ for $C$ where the circuit associated with each region is a $\left(\frac{n}{(100k)^{d-2}}, m, 2, k\right)$-circuit (either a $k$-CNF or a $k$-DNF). With probability at least $1-q$, $|\mathcal{P}| \leq s$ and the algorithm runs in time $\text{poly}(n) \cdot \lg \frac{1}{q} \cdot |C| \cdot s$ where $s \leq \frac{(2n)^{d-2}}{(100k)^{(d-1)(d-2)/2}} 2^{n - \frac{n}{(100k)^{d-2}} + (d-2)3^{-k}m}$.*

*Proof.* We will prove Lemma 3.4 by induction on $d$. If $d = 2$, output $((R=1, \rho=1), C)$ since $C$ is already a $k$-CNF or $k$-DNF and the bounds on $\mathcal{P}$ and the running time holds with probability 1.

If $d > 2$ we assume by induction that we can run this algorithm recursively on circuits of depth $d-1$ and that the recursive call will satisfy the properties of the lemma. First, run the Depth Reduction Algorithm (Lemma 3.2) on $(C, q = q/2)$ to get $\mathcal{P} = \{\mathcal{R}_i, C_i\}_i$ (where each $C_i$ is a $(\frac{n}{100k}, m, d-1, k)$-circuit). Then for each $i$, since $C_i$ has depth $d-1$, run this algorithm recursively on the $(C_i, q = q/2^{n+1})$ to get $\mathcal{P}_i = \{(\mathcal{R}_{i,j}, C_{i,j})\}_j$ (where each $C_{i,j}$ is a $(\frac{n}{(100k)^{d-2}}, m, 2, k)$-circuit). For each $j$, output $(\mathcal{R}_i \wedge \mathcal{R}_{i,j}, C_{i,j})$.

Say that $\mathcal{P}$ is good if $|\mathcal{P}| \leq \frac{2n}{100k} 2^{n(1-\frac{1}{100k}) + 3^{-k}m}$. By Lemma 3.2, $\mathcal{P}$ is good with probability at least $1 - q/2$. For each $i$, say that $\mathcal{P}_i$ is good if $|\mathcal{P}_i| \leq \frac{\left(\frac{2n}{100k}\right)^{d-3}}{(100k)^{(d-2)(d-3)/2}} 2^{\left(\frac{n}{100k}\right)(1-\frac{1}{(100k)^{d-3}}) + (d-3)3^{-k}m}$ (note that this is the result of a recursive call on a $(\frac{n}{100k}, m, d-1, k)$-circuit). By induction, each $\mathcal{P}_i$ is good independently with probability at least $1 - q/2^{n+1}$. By a union bound, $\mathbf{Pr}[\mathcal{P}_i \text{ is good for all } i \mid \mathcal{P} \text{ is good}] \geq 1 - q/2$, since if $\mathcal{P}$ is good then the number of $\mathcal{P}_i$s is much less than $2^n$. The probability that $\mathcal{P}$ and $\mathcal{P}_i$ for all $i$ are all good is at least $(1-q/2)(1-q/2) > 1-q$. In this case, the total number of outputs is at most

$$\left(\frac{2n}{100k} \cdot 2^{n(1-\frac{1}{100k})+3^{-k}m}\right) \left(\frac{\left(\frac{2n}{100k}\right)^{d-3}}{(100k)^{(d-2)(d-3)/2}} 2^{\left(\frac{n}{100k}\right)(1-\frac{1}{(100k)^{d-3}})+(d-3)3^{-k}m}\right)$$
$$= \frac{(2n)^{d-2}}{(100k)^{(d-1)(d-2)/2}} 2^{n(1-\frac{1}{(100k)^{d-2}})+(d-2)3^{-k}m}.$$

□

When we end up with a $k$-CNF or a $k$-DNF $C$ in a region defined by a $k$-CNF $R$ (and a restriction), we apply a random restriction on a $1 - 1/O(k)$ fraction of variables and then construct a decision tree for the pair $(C, R)$ and then argue that the total number of leaves in the decision tree isn't too big. Each leaf in the decision tree where $R \equiv 1$ corresponds to a restriction in the region $R$ where the value of the circuit is constant.

**Lemma 3.5** (Depth Two Algorithm). *Let $C$ be a $k$-CNF or $k$-DNF and $R$ be a $k$-CNF each in the same $n$ variables and let $0 < q \leq 1/2$ be a parameter. There exists a randomized algorithm which outputs a partitioning $\mathcal{P}$ for $C$ in the region $R$ where each region on $\mathcal{P}$ is defined by a restriction and the circuit associated with each region of $\mathcal{P}$ is either $0$ or $1$. With probability at least $1-q$, $|\mathcal{P}| \leq s$ and the algorithm runs in time $\text{poly}(n) \cdot \lg \frac{1}{q} \cdot (|C| + |R|) \cdot s$ where $s \leq 50 \cdot 2^{n\left(1-\frac{1}{30k}\right)}$.*

*Proof.* Let $p = \frac{1}{30k}$. Choose a set $U$ of $pn$ variables uniformly at random. For each restriction $\rho$ which leaves the variables in $U$ unset, construct the decision tree $T$ for $(C, R)|_\rho$. For each path $\rho'$ in $T$ ending at a leaf labeled $(b, 1)$, output $(\rho \wedge \rho', b)$.

The regions defined by each $\rho \wedge \rho'$ output partition the region $R$, and in each region $C$ has value $b$. All that remains is bounding the number pairs output. By Corollary 1.5, $\mathbf{Pr}_\rho[height(tree((C,R)|_\rho)) \geq s] \leq$ $3(13/30)^s$. We bound the expected total number of outputs by first summing over all $\rho$ and then summing over $s$ and using the fact that a decision tree of hight at most $s$ can have at most $2^s$ leaves.

$$\mathbf{E}_{U}[|\mathcal{P}|] \leq \mathbf{E}_{U}\left[\sum_{\rho} \text{number of leaves of } tree((C,R)|_\rho)\right]$$

$$= 2^{n\left(1-\frac{1}{30k}\right)} \mathbf{E}_{U,\rho}[\text{number of leaves of } tree((C,R)|_\rho)]$$

$$\leq 2^{n\left(1-\frac{1}{30k}\right)} \sum_{s=0}^{\frac{n}{30k}} 2^s 3(13/30)^s \leq 3 \cdot 2^{n\left(1-\frac{1}{30k}\right)} \sum_{s=0}^{\infty}(26/30)^s < 25 \cdot 2^{n\left(1-\frac{1}{30k}\right)}.$$

By Markov's inequality, with probability at least $1/2$, $|\mathcal{P}| \leq 50 \cdot 2^{n\left(1-\frac{1}{30k}\right)}$. We may repeat the algorithm $\lg \frac{1}{q}$ times in parallel with independent choices of $U$ and output the smallest partition. This increases the probability of success to $1 - q$. □

We compose the preceding algorithms to get our algorithm for $\mathbf{AC}^0$ circuits, formalized in the following lemma.

**Lemma 3.6.** *Let $C$ be a $(n,m,d)$-circuit. Let $k \geq 1$ be a parameter. There exists a randomized algorithm which outputs a partitioning $\mathcal{P}$ for $C$ where each region is defined by a restriction and the circuit associated with each region is either $0$ or $1$. With probability at least $1 - 2^{-n}$, $|\mathcal{P}| \leq s$ and the algorithm runs in time at most $poly(n) \cdot |C| \cdot s$ where $s \leq 50 \frac{(2n)^{d-2}}{(100k)^{(d-1)(d-2)/2}} 2^{n - \frac{3n}{(100k)^{d-1}} + (d-2)3^{-k}m + 4 \cdot 2^{-k}\max(m,n/k)}$.*

Theorem 1.2 follows straightforwardly from Lemma 3.6.

*Proof of Theorem 1.2.* Let $C$ be a $(n,m,d)$-circuit. If $m \geq 2^{\Omega(n)^{\frac{1}{d-1}}}$ then output all $2^n$ restrictions which set all of the variables. Otherwise, we may choose $k = \Theta\left(\max\left(\lg \frac{m}{n}, d \lg d\right)\right)$ such that

$$50 \frac{(2n)^{d-2}}{(100k)^{(d-1)(d-2)/2}} 2^{(d-2)3^{-k}m + 4 \cdot 2^{-k}\max(m,n/k)} \leq 2^{\frac{2n}{(100k)^{d-1}}}$$

and then use Lemma 3.6. In either case, the algorithm outputs at most $2^{n - \frac{n}{O(\lg \frac{m}{n} + d \lg d)^{d-1}} + O(1)}$ restrictions. □

For the sake of simplifying the calculations in the proof of Lemma 3.6, we first prove the special case where the circuit has bottom fan-in $k$.

**Lemma 3.7.** *Let $C$ be a $(n,m,d,k)$-circuit. There exists a randomized algorithm which outputs a partitioning $\mathcal{P}$ for $C$ where each region is defined by a restriction and the circuit associated with each region is either $0$ or $1$. With probability at least $1 - 2^{-2n}$, $|\mathcal{P}| \leq s$ and the algorithm runs in time at most $poly(n) \cdot |C| \cdot s$ where $s \leq 50 \frac{(2n)^{d-2}}{(100k)^{(d-1)(d-2)/2}} 2^{n - \frac{3n}{(100k)^{d-1}} + (d-2)3^{-k}m}$.*

*Proof.* Run the Repeated Depth Reduction Algorithm (Lemma 3.4) on $(C, q/2)$ to get a partition $\mathcal{P} = \{((R_i, \rho_i), C_i)\}_i$ for $C$. Each $C_i$ is either a $k$-CNF or a $k$-DNF in $\frac{n}{(100k)^{d-2}}$ variables. Run the Depth Two Algorithm (Lemma 3.5) on $(C_i, R_i, q = q/2^{n+1})$ to get a partition $\mathcal{P}_i = \{(\rho_{i,j}, b_{i,j})\}_j$ for $C_i$. Output $(\rho_i \wedge \rho_{i,j}, b_{i,j})$.

Say that $\mathcal{P}$ is good if $|\mathcal{P}| \leq \frac{(2n)^{d-2}}{(100k)^{(d-1)(d-2)/2}} 2^{n(1-\frac{1}{(100k)^{d-2}})+(d-2)3^{-k}m}$. By Lemma 3.4, $\mathbf{Pr}\left[\mathcal{P} \text{ is good}\right] \geq 1-q/2$. For each $i$, say that $\mathcal{P}_i$ is good if $|\mathcal{P}_i| \leq 50 \cdot 2^{\frac{n}{(100k)^{d-2}}(1-\frac{1}{30k})}$ (note that the Depth Two Algorithm is run on circuits in $\frac{n}{(100k)^{d-2}}$ variables). By Lemma 3.5, $\mathbf{Pr}\left[\mathcal{P}_i \text{ is good}\right] \geq 1 - q/2^{n+1}$. By a union bound, $\mathbf{Pr}\left[\mathcal{P}_i \text{ is good for all } i \mid \mathcal{P} \text{ is good}\right] \geq 1-q/2$, since if $\mathcal{P}$ is good then the number of $\mathcal{P}_i$s is much less than $2^n$. The probability that $\mathcal{P}$ and $\mathcal{P}_i$ are all good is at least $(1-q/2)(1-q/2) > 1-q$. In this case, the total number of outputs is at most

$$\left(\frac{(2n)^{d-2}}{(100k)^{(d-1)(d-2)/2}} 2^{n(1-\frac{1}{(100k)^{d-2}})+(d-2)3^{-k}m}\right) \left(50 \cdot 2^{\frac{n}{(100k)^{d-2}}(1-\frac{1}{30k})}\right)$$

$$\leq 50 \frac{(2n)^{d-2}}{(100k)^{(d-1)(d-2)/2}} 2^{n(1-\frac{3}{(100k)^{d-1}})+(d-2)3^{-k}m}.$$

□

*Proof of Lemma 3.6.* Run the Bottom Fan-in Reduction Algorithm (Lemma 3.1) on $(C, k)$ to get $\mathcal{P} = \cup_{0 \leq f \leq n/k} \mathcal{P}_f$ where $\mathcal{P}_f = \{((R_{f,i}, \rho_{f,i}), C_{f,i})\}_i$. For each $f$ and $i$, the circuit $C_{f,i}$ is an $(n-fk, m, d, k)$-circuit. Run the algorithm of Lemma 3.7 on $(C_{f,i}, q = 2^{-2n})$ to get a partition $\mathcal{P}_{f,i} = \{(\rho_{f,i,j}), b_{f,i,j}\}_j$ for $C_{f,i}$. Output $(\rho_{f,i} \wedge \rho_{f,i,j}, b_{f,i,j})$.

By Lemma 3.1, the sets $\mathcal{P}_f$ are disjoint and for each $f$, $|\mathcal{P}_f| \leq \binom{m+f}{f}$ and the circuits associated with each region in $\mathcal{P}_f$ are $(n-fk, m, d, k)$-circuits. For each $f$ and $i$, say that $\mathcal{P}_{i,f}$ is good if $|\mathcal{P}_{f,i}| \leq 50 \frac{(2n)^{d-2}}{(100k)^{(d-1)(d-2)/2}} 2^{(n-fk)(1-\frac{3}{(100k)^{d-1}})+(d-2)3^{-k}m}$. By Lemma 3.7, $\mathbf{Pr}\left[\mathcal{P}_{f,i} \text{ is good}\right] \geq 1 - 2^{-2n}$, and by a union bound over the at most $2^n$ pairs $f, i$, all of the sets $\mathcal{P}_{f,i}$ are simultaneously good with probability at least $1 - 2^{-n}$. In this case, the total number of outputs is at most

$$\sum_{f=0}^{n/k} \binom{m+f}{f} 50 \frac{(2n)^{d-2}}{(100k)^{(d-1)(d-2)/2}} 2^{(n-fk)(1-\frac{3}{(100k)^{d-1}})+(d-2)3^{-k}m}$$

$$= 50 \frac{(2n)^{d-2}}{(100k)^{(d-1)(d-2)/2}} 2^{n(1-\frac{3}{(100k)^{d-1}})+(d-2)3^{-k}m} \sum_{f=0}^{n/k} \binom{m+f}{f} 2^{-fk(1-\frac{3}{(100k)^{d-1}})+(d-2)3^{-k}m}$$

$$\leq 50 \frac{(2n)^{d-2}}{(100k)^{(d-1)(d-2)/2}} 2^{n(1-\frac{3}{(100k)^{d-1}})+(d-2)3^{-k}m+4\cdot 2^{-k}\max(m,n/k)}$$

since

$$\sum_{f=0}^{n/k} \binom{m+f}{f} 2^{-fk(1-\frac{3}{(100k)^{d-1}})} \leq \sum_{f=0}^{m+n/k} \binom{m+n/k}{f} 2^{-fk} = \left(1 + 2^{-k}\right)^{m+n/k}$$

$$\leq 2^{(\lg e)2^{-k}(m+n/k)} \leq 2^{4\cdot 2^{-k}\max(m,n/k)}$$

□

## 3.1 Switching Algorithm

**Lemma** (Lemma 3.3, restated). *Let $\Phi = (\phi_1, \ldots, \phi_m)$ be a sequence of $k$-CNFs in $n$ variables and let $0 < q \leq 1/2$ be a parameter. There exists a randomized algorithm which takes $\Phi$ as input and outputs a partitioning $\mathcal{P}$ for $\Phi$ where the circuits in each region of $\mathcal{P}$ are $k$-DNFs in at most $\frac{n}{100k}$ variables. With probability at least $1 - q$, $|\mathcal{P}| \leq s$ and the algorithm runs in time at most $\mathrm{poly}(n) \cdot \lg \frac{1}{q} \cdot |\Phi| \cdot s$ where $s \leq \frac{2n}{100k} 2^{n-\frac{n}{100k}+3^{-k}m}$.*

Let $\phi$ be a $k$-CNF. Consider the decision tree $tree(\phi)$ for $\phi$. If the height of $tree(\phi)$ is at most $k'$, then we can construct a $k'$-DNF $\phi'$ equivalent to $\phi$ by taking the disjunction of the terms corresponding to paths in $tree(\phi)$ labeled 1.

However, in general, $k'$ will be much larger than $k$. In this case, we split the paths in $tree(\phi)$ into two categories: "short paths" of length at most $k$, and "long paths" of length greater than $k$. Since any assignment to a set variables is consistent with exactly one path in any decision tree on those variables, we can partition $\{0,1\}^n$ by partitioning the paths. We will construct a $k$-CNF $\mathcal{T}(\phi, k)$ which will define the region corresponding to the set of short paths. Rather than defining a single region for the set of long paths, we further partition the space and define a separate region for each long path (each long path viewed as a restriction defines a region).

Formally, for any $k$-CNF $\phi$ and any $k \geq 1$, let $\Sigma' = \{\sigma_1', \ldots, \sigma_{\ell'}'\}$ be the set of paths of length greater than $k$ in $tree(\phi)$. Let $\Sigma = \{\sigma_1, \ldots, \sigma_\ell\}$ be the set of paths of length $k$ in $tree(\phi)$ that do not end at a leaf (equivalently, $\Sigma$ consists of the paths in $\Sigma'$ truncated after $k$ variables). Define $\mathcal{T}(\phi, k) = \neg \sigma_1 \wedge \neg \sigma_2 \wedge \cdots \wedge \neg \sigma_\ell$ where each $\sigma_i$ is viewed as the conjunction of the literals along the path. Note that $\mathcal{T}(\phi, k)$ is a $k$-CNF. An assignment is in the region corresponding to short paths if and only if it is not consistent with any $\sigma_i'$. Since the paths $\sigma_i'$ form complete decision trees after the first $k$ variables along each path, an assignment is not consistent with any $\sigma_i'$ if and only if it is not consistent with any $\sigma_i$, and therefore if and only if it satisfies $\mathcal{T}(\phi, k)$.

The algorithm will branch on the region $\mathcal{T}(\phi, k)$ and on the regions corresponding to long paths. In the region $\mathcal{T}(\phi, k)$, we can do as we did before and convert $\phi$ into an equivalent $k$-DNF by only considering the short paths ending with 1. Formally, for any $k$-CNF $\phi$ and any $k \geq 1$, let $\mathcal{S}(\phi, k) = \tau_1 \vee \tau_2 \vee \cdots \vee \tau_\ell$ where $\tau_1, \ldots, \tau_\ell$ are the paths of the decision tree for $\phi$ of length at most $k$ with label 1 where each $\tau_i$ is viewed as a conjunction of the literals along the path. $\mathcal{S}(\phi, k)$ is a $k$-DNF and is equivalent to $\phi$ in the region $\mathcal{T}(\phi, k)$. In the regions correspond to long paths, $\phi$ is constant and therefore trivially a $k$-DNF.

We repeat this process for each $k$-CNF $\phi_1, \ldots, \phi_m$ in order, each time recursively partitioning the current set of regions.

The problem with this approach is that we may well branch more than $2^n$ times between branching on $\mathcal{T}(\phi, k)$ and branching on each long path. We solve this problem by first applying a random restriction which leaves a small constant fraction of the variables unset, and then do depth reduction as described above. In this case, we can bound the probability that the total number of variables set along long paths will be too large using Lemma 1.4.

*Proof.* Let $p = \frac{1}{100k}$. Choose a set $U$ of $pn$ variable to leave unset uniformly at random. Branch on each restriction $\rho_0$ which leaves the variables in $U$ unset. Let $\rho = \rho_0$ and let $R$ denote an empty (true) $k$-CNF. For each $\phi_i$ in order, branch on $\mathcal{T}(\phi_i|_\rho, k)$.

In the region $\rho \wedge \mathcal{T}(\phi_i|_\rho, k)$, $\phi_i$ is equivalent to $\mathcal{S}(\phi_i|_\rho, k)$ so set $\psi_i = \mathcal{S}(\phi_i|_\rho, k)$. Let $R = R \wedge \mathcal{T}(\phi_i|_\rho, k)$. In this branch, say that $\phi_i$ is "not targeted."

When $\mathcal{T}(\phi_i|_\rho, k)$ is false, we further branch on each path $\rho'$ in $tree(\phi_i|_\rho)$ of length greater than $k$. Let $\rho = \rho \wedge \rho'$ and let $\psi_i = b'$ where $b'$ is the label at the end of the path $\rho'$. In each of these branches, say that $\phi_i$ is "targeted."

Once we have branched in this fashion for each $\phi_i$, output the resulting $(\mathcal{R} = (R, \rho), \Psi = (\psi_1, \ldots, \psi_m))$.

This algorithm naturally defines a computation tree: First branch on each restriction $\rho_0$, then for each $\phi_i$ in order branch on whether $\phi_i$ is targeted, and if $\phi_i$ is targeted branch on each long path. Each leaf of this computation tree correspond to a region in partition output. Define the "type" of each leaf as the sequence of targeted $k$-CNFs along the path to the leaf. We will group the leaves of the computation tree both by type and by parent restriction $\rho_0$ in order to bound the expected number of leaves.

For any $\rho_0$ and any type (sequence of targeted $k$-CNFs) $T$, let $G_{\rho_0,T}$ denote the set of leaves of type $T$ with parent restriction $\rho_0$. Consider the decision tree $tree(T|_{\rho_0})$. Let $P_{\rho_0,T}$ denote the set of paths in $tree(T|_{\rho_0})$ where each $k$-CNF in $T$ contributes at least $k+1$ variables to the path. The set of paths $P_{\rho_0,T}$ correspond exactly to the leaves in $G_{\rho_0,T}$: The path in $P_{\rho_0,T}$ corresponds to which branch $\rho'$ is taken at each targeted $k$-CNF (each $\rho'$ sets at least $k+1$ variables since only paths of length $> k$ are considered when branching).

By Lemma 1.4, $\mathbf{Pr}_{U,\rho_0}[tree(T|_{\rho_0})$ has a path of length $\geq s$ where each $k$-CNF in $T$ contributes at least one variable$] \leq (13/100)^s$, which gives $\mathbf{E}_{U,\rho_0}[|P_{\rho_0,T}|] \leq \sum_{s=0}^{pn} 2^s (13/100)^s$ (any decision tree of height $s$ can have at most $2^s$ leaves). Since we only consider paths in $tree(T|_{\rho_0})$ where each $\phi_i \in T$ contributes at least $k+1$ variables, any such path must have length at least $|T|(k+1)$. We bound the expected number of outputs by summing over restrictions $\rho_0$, then path lengths $s$ and then sets of targeted $k$-CNFs $T$ of size at most $s/(k+1)$.

$$\mathbf{E}_U[|\mathcal{P}|] = \sum_{\rho_0} \sum_{T \subseteq \{\phi_1,\ldots,\phi_m\}} \mathbf{E}_U[|P_{\rho_0,T}|] = 2^{n-pn} \sum_{T \subseteq \{\phi_1,\ldots,\phi_m\}} \mathbf{E}_{U,\rho_0}[|P_{\rho_0,T}|]$$

$$\leq 2^{n-pn} \sum_{T \subseteq \{\phi_1,\ldots,\phi_m\}} \sum_{s=0}^{pn} 2^s \mathbf{Pr}_{U,\rho_0}\begin{bmatrix} tree(T|_{\rho_0}) \text{ has a path of length } \geq s \\ \text{where each } \phi_i \in T \text{ contributes} > k \text{ vari-} \\ \text{ables} \end{bmatrix}$$

$$= 2^{n-pn} \sum_{s=0}^{pn} \sum_{\substack{T \subseteq \{\phi_1,\ldots,\phi_m\} \\ |T| \leq s/(k+1)}} 2^s \mathbf{Pr}_{U,\rho_0}\begin{bmatrix} tree(T|_{\rho_0}) \text{ has a path of length } \geq s \\ \text{where each } \phi_i \in T \text{ contributes} > k \text{ vari-} \\ \text{ables} \end{bmatrix}$$

$$\leq 2^{n-pn} \sum_{s=0}^{pn} \sum_{t=0}^{\lfloor s/k \rfloor} \binom{m}{t} 2^s (13/100)^s \leq 2^{n-pn} \sum_{s=0}^{pn} \left\lfloor \frac{s}{k} \right\rfloor \binom{m}{\lfloor s/k \rfloor} (26/100)^s$$

grouping terms with the same value of $\lfloor s/k \rfloor$ and replacing $\lfloor s/k \rfloor$ with $s'$

$$\leq 2^{n-pn} \sum_{s'=0}^{pn/k} ks' \binom{m}{s'} (26/100)^{ks'} \leq pn 2^{n-pn} \sum_{s'=0}^{m} \binom{m}{s'} (26/100)^{ks'}$$

$$= \frac{n}{100k} 2^{n-pn} (1 + (26/100)^k)^m \leq \frac{n}{100k} 2^{n-pn+(\lg e)(26/100)^k m}.$$

With probability at least $1/2$, $|\mathcal{P}| \leq \frac{2n}{100k} \cdot 2^{n-\frac{n}{100k}+3^{-k}m}$. We may repeat the algorithm $\lg \frac{1}{q}$ times in parallel with independent choices of $U$ and output the smallest partition. This increases the probability of success to $1 - q$. □

## 3.2 Extended Switching Lemma

**Lemma** (Lemma 1.4, Extended Switching Lemma, restated). *Let $\phi_1, \ldots, \phi_m$ be a sequence of $k$-CNFs and/or $k$-DNFs in the same $n$ variables. For any $p \leq 1/13$, let $\rho$ be a random restriction which leaves $pn$ variables unset. The probability that the decision tree for $(\phi_1, \ldots, \phi_m)|_\rho$ has a path of length $\geq s$ where each $\phi_i$ contributes at least one node to the path is at most $(13pk)^s$.*

Our switching lemma is based on Beame's proof [1] of Razborov's Switching Lemma [16] which in turn is based on Håstad's Switching Lemma [5]. The idea is to encode a restriction and a corresponding "bad" path in the resulting decision tree using a different restriction which sets additional variables and a few extra bits. The total size needed for this encoding will be much less than the size needed to encode the original restriction and this ratio will give the probability bound.

*Proof.* Let $P'$ be a path in $tree((\phi_1, \ldots, \phi_m)|_\rho)$ of length at least $t$. Let $P$ denote the prefix of $P'$ of length exactly $t$. Let $m'$ denote the index of the last formula $\phi_{m'}$ which contributes a variable to $P$. Let $x_1, \ldots, x_t$ denote the variables along $P$ and let $p_1, \ldots, p_t$ denote the values that $P$ assigns to $x_1, \ldots, x_t$. Let $C_1, \ldots, C_t$ and $F_1, \ldots, F_t$ denote the clauses and formulae respectively which contribute $x_1, \ldots, x_t$ (some $C_i$s and $F_i$s may refer to the same clauses or formulae if they contribute more than one variable to $P$). Let $index_1, \ldots, index_t$ denote the indices of $x_1, \ldots, x_t$ in the clauses $C_1, \ldots, C_t$. Let $last_i, 1 \leq i \leq t$ be 2 if $x_i$ is the last variable contributed by $F_i$ along $P$; be 1 if $x_i$ is the last variable contributed by $C_i$ (but not by $F_i$) along $P$; and be 0 otherwise.

Let $\sigma = \sigma_1 \cdots \sigma_{s+1}$ where $\sigma_i$ is a restriction where $\sigma_i(x_i) = 0$ if $x_i$ appears positively in $C_i$ and $\sigma_i(x_i) = 1$ otherwise, and $\sigma_i(y) = *$ for all $y \neq x_i$ ($\sigma$ is constructed to *not satisfy* the clauses $C_1, \ldots, C_{s+1}$). Let $\pi_i$ be the restriction where $\pi_i(x_i) = p_i$ and $\pi_i(y) = *$ for $y \neq x_i$. Note that $P = \pi_1 \cdots \pi_{s+1}$

We map $(\rho, P)$ to $\rho' = \rho\sigma \in \mathcal{R}_n^{pn-t}, \vec{index} = (index_1, \ldots, index_t) \in [k]^t, \vec{last} = (last_1, \ldots, last_t) \in [3]^t$ and $\vec{p} = (p_1, \ldots, p_t) \in [2]^t$.

We must now show that we can decode $(\rho, P)$ from $\rho', \vec{index}, \vec{last}$ and $\vec{p}$. Let $\rho_i = \rho\pi_1 \cdots \pi_i \sigma_{i+1} \cdots \sigma_{s+1}$, for $0 \leq i \leq s+1$. Note that $\rho_0 = \rho'$ and $\rho_{s+1} = \rho P$.

We will show that for any $i < s+1$ given $C_i, F_i$ and $\rho_i$, we can decode $\pi_{i+1}$ and therefore $\rho_{i+1}$. Then by induction, given $\rho'$ we can decode $\rho$ and $P$. Let $last_0 = 1$ and let $F_0 = \phi_1$. First we identify $F_{i+1}$. If $last_i = 2$ then $F_{i+1} = \phi_{q+1}$ where $q$ is the index such that $F_i = \phi_q$, and otherwise $F_{i+1} = F_i$. If $last_i = 0$ then we know $C_{i+1} = C_i$, otherwise we claim that $C_{i+1}$ is the first clause not satisfied by $\rho_i$. Once we identify $C_{i+1}$, then $index_{i+1}$ is the index of $x_{i+1}$ in this clause and we get $\pi_i$ using $p_i$. All that remains is to prove the claim that when $last_i = 1$ then $C_{i+1}$ is first clause in $F$ not satisfied by $\rho_i$.

When $x_{i+1}$ is queried along the path $P$ when constructing the decision tree for $F|_\rho$, $C_{i+1}$ is the first clause not satisfied by $\rho\pi_1 \cdots \pi_i$ (otherwise a variable in an earlier clause would have been queried instead). Since setting more variables in a restriction cannot change a clause from satisfied to not satisfied, all the clauses in $F$ before $C_i$ are satisfied by $\rho_i$. All that remains is to show that $\sigma_{i+1} \cdots \sigma_{s+1}$ does not satisfy $C_{i+1}$. Let $j$ be the largest index such that $C_j = C_{i+1}$, or equivalently $x_j$ is the last variable from $C_{i+1}$ along $P$. By the construction of $\sigma_{i+1}, \ldots, \sigma_j$, $\sigma_{i+1} \cdots \sigma_j$ does not satisfy $C_{i+1}$ and either $j = t$ or $\rho\pi_1 \cdots \pi_i \sigma_{i+1} \cdots \sigma_j$ sets all of the variables in $C_{i+1}$ (because of the way we construct decision trees). In either case, no $\sigma_\ell, \ell > j$ can satisfy $C_{i+1}$ so we conclude that $\rho_i$ does not satisfy $C_{i+1}$.

All that remains is to calculate the probability over the choice of $\rho$ that the path $P$ exists. We bound this probability by the size of the encoding when $P$ exists over the size of the encoding of $\rho$.

$$\mathbf{Pr}_{\rho}\begin{bmatrix}tree((\phi_1,\ldots,\phi_m)|_{\rho}) \text{ has a} \\ \text{path of length } \geq s \text{ where} \\ \text{each } \phi_i \text{ contributes at least} \\ \text{one node to the path}\end{bmatrix} \leq \frac{|\mathcal{R}_n^{pn-t} \times [k]^t \times [3]^t \times [2]^t|}{|\mathcal{R}_n^{pn}|} = \frac{\binom{n}{pn-t}2^{n-pn+t}(6k)^t}{\binom{n}{pn}2^{n-pn}}$$

$$= \frac{(pn)!(n-pn)!}{(pn-t)!(n-pn+t)!}(12k)^t$$

$$= \left(\frac{pn}{n-pn+t}\right)\left(\frac{pn-1}{n-pn+t-1}\right)\cdots\left(\frac{pn-t+1}{n-pn+1}\right)(12k)^t$$

$$\leq \left(\frac{pn}{n-pn+t}\right)^t(12k)^t \leq \left(\frac{12pk}{1-p}\right)^t \leq (13pk)^t.$$

$\square$

## 4 Correlation & Lower Bounds

Let $C$ be a depth $d$ size $m$ $\mathbf{AC}^0$ circuit. Theorem 1.2 implies that there exists a set of at most $2^{n-n/O(\lg m + d\lg d)^{d-1}}$ restrictions which partition $\{0,1\}^n$ and make $C$ constant.

Consider the parity functions. Any partition of $\{0,1\}^n$ which makes parity constant requires $2^n$ restrictions. Setting $2^{n-n/O(\lg m + d\lg d)^{d-1}} \geq 2^n$ and solving for $m$ in terms of $d$ and vice versa we get the following bounds, which match the optimal bounds proved by Håstad [5]:

- Any size $poly(n)$ $\mathbf{AC}^0$ circuit which computes parity requires depth at least $\frac{\lg n}{\lg \lg n} - O\left(\frac{\lg n}{\lg^2 \lg n}\right)$.

- Any depth $d$ circuit which computes parity requires size at least $2^{\Omega\left(n^{\frac{1}{d-1}}\right)}$.

**Definition 4.1.** Let $\mathcal{C}$ denote a class of circuits. Let $f : \{0,1\}^n \to \{0,1\}$ be a function. The *correlation* of $f$ with circuits from $\mathcal{C}$ is

$$\max_{C\in\mathcal{C}}\left(\mathbf{Pr}_{x\in\{0,1\}^n}[C(x)=f(x)] - \mathbf{Pr}_{x\in\{0,1\}^n}[C(x)\neq f(x)]\right) = \max_{C\in\mathcal{C}}\mathbf{E}_{x\in\{0,1\}^n}\left[(-1)^{C(x)}(-1)^{f(x)}\right]$$

*Proof of Theorem 1.3.* Let $\mathcal{C}$ be the class of depth $d$ size $cn$ $\mathbf{AC}^0$ circuits. We will bound the correlation of $\mathcal{C}$ and the parity function. Let $C$ be an element of $\mathcal{C}$ and consider the partition produce by Theorem 1.2. Each restrictions in this partition which sets fewer than $n$ variables contributes 0 to the correlation with parity. Each restriction which sets all $n$ variables contributes at most $2^{-n}$ to the correlation with parity. Thus the correlation of $\mathcal{C}$ with parity is at most $2^{-\mu_{c,d}n} = 2^{-n/O(\lg c + d\lg d)^{d-1}}$. $\square$

We give a construction of a family of $(n,cn,d)$-circuits that both require $2^{n-n/\Omega(\lg c)^{d-1}}$ regions and have correlation $2^{-n/\Omega(\lg c)^{d-1}}$ with parity. Thus, our algorithm and the implied correlation bounds with parity are close to optimal. This construction is generally considered folklore, and a more detailed explanation is given in [8].

Parity on $\ell$ inputs can be computed by depth $d$ size $O(\ell 2^{\ell^{\frac{1}{d-1}}})$ circuits with either an $\wedge$ or an $\vee$ output gate. Construct a circuit on $n$ inputs by grouping the inputs into $\frac{n}{\Theta(\lg c)^{d-1}}$ groups of $\ell = \Theta\left(\lg c\right)^{d-1}$ inputs. Construct depth $d$ circuits computing the parity of each group each with an $\wedge$ output gate. Take the conjunction of all of these circuits. This gives a $(n,cn,d)$-circuit.

Each input which sets the circuit to 1 correctly computes parity (assume that the number of groups of inputs is odd). At most half the remaining inputs compute parity correctly. The fraction of inputs which cause the circuit to output one, and therefore the correlation with parity is $2^{-\frac{n}{\Theta(\lg c)^{d-1}}}$. The number of inputs which cause the circuit to output one is $2^{n-\frac{n}{\Theta(\lg c)^{d-1}}}$. Since flipping any bit on any of these inputs will change the output to 0, each input which causes the circuit to output 1 must have its own region, giving a $2^{n-\frac{n}{\Omega(\lg c)^{d-1}}}$ lower bound on the number of regions.

Finally, we will sketch how a small but significant improvement in our algorithm combined with the results of Williams [20, 21] would show new circuit lower bounds. If the satisfiability of size $m$, depth $d$ **AC**$^0$ circuits in $n$ variables can be decided in time $2^{n(1-\frac{1}{O(\lg m)^{o(d)}})}$, then **NEXP** $\not\subseteq$ **NC**$^1$. Note that any such algorithm *cannot* do as we did and enumerate satisfying restrictions since there can be too many.

Any **NC**$^1$ circuit $C$ with $n$ inputs can be converted to an equivalent depth $d$ size $2^{n^{O(\frac{1}{d-1})}}$ **AC**$^0$ circuit, by the following construction which seems to be considered folklore. Let $D = O(\lg n)$ denote the depth of $C$. Divide the layers of $C$ into $d-1$ groups of $\frac{D}{d-1}$ consecutive layers. Replace each group of layers by trivial depth two, size $poly(n)2^{2^{\frac{D}{d-1}}}$ circuits, alternating between ANDs of ORs and ORs of ANDs, and then combine layers of the same type of gates. Since $D = O(\lg n)$, the result is a depth $d$ size $2^{n^{O(\frac{1}{d-1})}}$ **AC**$^0$ circuit.

Williams shows that for any "reasonable" class of circuits $\mathcal{C}$ (**AC**$^0$, **ACC**, **NC**$^1$, **P**/*poly*, etc.), if the satisfiability of circuits in $\mathcal{C}$ with $n$ inputs is in co-nondeterministic time $2^{n-n^\epsilon}$ for some constant $\epsilon > 0$ then **NEXP** $\not\subseteq \mathcal{C}$. Running a (hypothetical) $2^{n(1-\frac{1}{O(\lg m)^{o(d)}})}$ satisfiability algorithm on the size $2^{n^{O(\frac{1}{d-1})}}$ **AC**$^0$ circuit resulting from a **NC**$^1$ circuit would give a sufficiently fast satisfiability algorithm for Williams' result to show **NEXP** $\not\subseteq$ **NC**$^1$.

## 5  Acknowledgements

The first author would like to thank Srikanth Srinivasan and Paul Beame for raising the question of correlation of $AC^0$ with parity, and for many discussions of related topics. We are grateful to Valentine Kabanets for providing very useful feedback on an earlier draft of this paper, and to Daniel Lokshtanov for feedback and discussions on some of the intermediate ideas.

# References

[1] Paul Beame. A switching lemma primer. Technical Report, Department of Computer Science and Engineering, University of Washington, 1994.

[2] Paul Beame, Russell Impagliazzo, and Srikanth Srinivasan. Approximating ac0 circuits by 'small' height decision trees. manuscript, March 2011.

[3] Chris Calabro, Russell Impagliazzo, and Ramamohan Paturi. A duality between clause width and clause density for SAT. In *CCC '06: Proceedings of the 21st Annual IEEE Conference on Computational Complexity*, pages 252–260, Washington, DC, USA, 2006. IEEE Computer Society.

[4] Chris Calabro, Russell Impagliazzo, and Ramamohan Paturi. The complexity of satisfiability of small depth circuits. In *Parameterized and Exact Computation: 4th International Workshop, IWPEC 2009, Copenhagen, Denmark, September 10-11, 2009, Revised Selected Papers*, pages 75–85, Berlin, Heidelberg, 2009. Springer-Verlag.

[5] J. Håstad. Almost optimal lower bounds for small depth circuits. In *Proceedings of the eighteenth annual ACM symposium on Theory of computing*, STOC '86, pages 6–20, New York, NY, USA, 1986. ACM.

[6] Johan Håstad. On parity. Unpublished Manuscript, 2011.

[7] Johan Håstad, Stasys Jukna, and Pavel Pudlák. Top-down lower bounds for depth-three circuits. *Computational Complexity*, 5(2):99–112, 1995.

[8] Johan Torkel Håstad. *Computational limitations for small-depth circuits*. MIT Press, Cambridge, MA, USA, 1987.

[9] Timon Hertli. 3-SAT Faster and Simpler — Unique-SAT Bounds for PPSZ Hold in General. 2011.

[10] Michael L. Littman, Toniann Pitassi, and Russell Impagliazzo. New and old algorithms for belief net inference and counting satisfying assignments. Unpublished Manuscript, 2001.

[11] Sharad Malik and Lintao Zhang. Boolean satisfiability from theoretical hardness to practical success. *Commun. ACM*, 52:76–82, August 2009.

[12] B. Monien and E. Speckenmeyer. Solving satisfiability in less than $2^n$ steps. *Discrete Appl. Math.*, 10:287–295, March 1985.

[13] Robin Moser and Domink Scheder. A Full Derandomization of Schöning's $k$-SAT Algorithm. In *Proceedings of the Forty-Fourth Annual ACM Symposium on Theory of Computing*, 2011.

[14] R. Paturi, P. Pudlák, M.E. Saks, and F. Zane. An improved exponential-time algorithm for $k$-SAT. *Journal of the ACM*, 52(3):337–364, May 2005. Preliminary version in *39th Annual IEEE Symposium on Foundations of Computer Science*, pages 628–637, 1998.

[15] R. Paturi, P. Pudlák, and F. Zane. Satisfiability coding lemma. *Chicago Journal of Theoretical Computer Science*, 1999. (preliminary version in FOCS'97).

[16] Alexander A. Razborov. Bounded arithmetic and lower bounds in boolean complexity. In *Feasible Mathematics II*, pages 344–386. Birkhauser, 1993.

[17] Rahul Santhanam. Fighting perebor: New and improved algorithms for formula and qbf satisfiability. In *Proceedings of the 2010 IEEE 51st Annual Symposium on Foundations of Computer Science*, FOCS '10, pages 183–192, Washington, DC, USA, 2010. IEEE Computer Society.

[18] U. Schöning. A probabilistic algorithm for $k$-SAT and constraint satisfaction problems. In *FOCS*, pages 410–414, 1999.

[19] R. Schuler. An algorithm for the satisfiability problem of formulas in conjunctive normal form. *Journal of Algorithms*, 54(1):40–44, 2005.

[20] Ryan Williams. Improving exhaustive search implies superpolynomial lower bounds. In *Proceedings of the 42nd ACM symposium on Theory of computing*, STOC '10, pages 231–240, New York, NY, USA, 2010. ACM.

[21] Ryan Williams. Non-Uniform ACC Circuit Lower Bounds. In *Proceedings of the Twenty-Sixth Annual IEEE Conference on Computational Complexity*, 2011.

[22] Andrew C-C. Yao. Separating the polynomial-time hierarchy by oracles. In *Proc. 26th annual symposium on Foundations of computer science*, pages 1–10, Piscataway, NJ, USA, 1985. IEEE Press.